\documentclass[pra,aps,superscriptaddress,twocolumn,amssymb,showpacs]{revtex4-1}

\usepackage{color}
\usepackage{graphicx}
\usepackage{longtable}
\usepackage{epsfig}
\usepackage{dcolumn}
\usepackage{bm}
\usepackage{amssymb}

\begin{document}

\title{Exotic Spin Order due to Orbital Fluctuations}

\author{     Wojciech Brzezicki}
\affiliation{Marian Smoluchowski Institute of Physics, Jagellonian
             University, Reymonta 4, PL-30059 Krak\'ow, Poland }

\author{     Jacek Dziarmaga}
\affiliation{Marian Smoluchowski Institute of Physics, Jagellonian
             University, Reymonta 4, PL-30059 Krak\'ow, Poland }

\author {    Andrzej M. Ole\'{s} }
\affiliation{Marian Smoluchowski Institute of Physics, Jagellonian
             University, Reymonta 4, PL-30059 Krak\'ow, Poland }
\affiliation{Max-Planck-Institut f\"ur Festk\"orperforschung,
             Heisenbergstrasse 1, D-70569 Stuttgart, Germany }

\date{31 January, 2014}

\begin{abstract}
We investigate the phase diagrams of the spin-orbital $d^9$
Kugel-Khomskii model for increasing system dimensionality: from the 
square lattice monolayer, via the bilayer to the cubic lattice. In 
each case we find strong competition between different types of spin 
and orbital order, with entangled spin-orbital phases at the 
crossover from antiferromagnetic to ferromagnetic correlations in
the intermediate regime of Hund's exchange. These phases have 
various types of exotic spin order and are stabilized by effective 
interactions of longer range which follow from enhanced spin-orbital 
fluctuations. We find that orbital order is in general more robust 
and spin order melts first under increasing temperature, as observed 
in several experiments for spin-orbital systems.
\end{abstract}


\maketitle

\section{Introduction}
\label{sec:intro}

The spin-orbital physics has started over 50 years ago when Kugel and 
Khomskii introduced the superexchange model for degenerate $e_g$ 
orbitals in KCuF$_3$ \cite{Kug73}, called the Kugel-Khomskii (KK) 
model. In cases when degenerate orbitals are partly filled both spins 
and orbitals have to be treated as quantum variables that are 
{\it a priori\/} strongly coupled to each other \cite{Kug82}. 
A similar situation occurs in a number of compounds with active orbital 
degrees of freedom, where strong on-site Coulomb interaction $U$ 
localizes charge carries (electrons or holes) and gives rise to 
spin-orbital superexchange \cite{Hfm,Kha05,Ole05,Ole09,Ole12}. 
A principal difficulty in such systems follows from enhanced quantum 
fluctuations \cite{Fei97,Kha97} which may destabilize long-range 
magnetic order  and could lead either to short-range 
spin-orbital correlations or to new quantum phases. Interplay between 
spin and orbital interactions leads to various types of magnetic order 
which coexist with particular orbital order, as in colossal 
magnetoresistance manganites \cite{Tok06}, or in the vanadium 
perovskites \cite{Fuj10}. The theoretical approaches are rare and 
notoriously difficult \cite{Goo06} --- while in 1D systems an electron 
can break into a spinon and an orbiton \cite{Woh11,Woh13}, explicit 
treatment of entangled spin-orbital states is required both in some 
model systems \cite{Ole06,Che07,Nor08,Cha11,Her11,You12,Lun12}, and in 
realistic models, for instance to describe the physical properties of 
the vanadium perovskites \cite{Ulr03,Hor03,Sir08,Hor08}. Therefore, the 
phase diagrams of such model systems are very challenging and are the 
subject of active research. 

Another route which makes the studies of spin-orbital physics of high 
interest is frustrated magnetism. Frustrated spin models are known to 
exhibit very interesting properties and have frequently exotic ground 
states \cite{Nor09}. In systems with active orbital degrees of freedom 
such states may arise from intrinsic frustration of orbital interactions 
which, unlike the spin ones with SU(2) symmetry, are directional both  
in the $e_g$ orbital models \cite{vdB99,vdB04,Ryn10} and 
in the compass model \cite{Mil05,Cin10,Tro10,Brz10,BrzDa} --- 
they contain terms which compete with one another. Studies of such 
models require more sophisticated approaches than the single-site
mean field (MF) approximation or linear spin-wave expansion. Exact 
solutions are possible only for some one-dimensional spin-orbital
models \cite{Fri99,Ole07,Kum13,Brz14} --- they also highlight the 
importance of quantum effects beyond simple classical approaches. 

Coming back to the KK model, it explains the origin of the orbital 
order in KCuF$_{3}$ which is responsible for the onset of the $A$-type 
antiferromagnetic ($A$-AF) order at low temperature \cite{Ole05},
and the quasi-1D AF Heisenberg structure with spinon excitations 
\cite{Ten93,Dei08,Lak13} at high temperature. 
In spite of strong interplay between the spin and orbital degrees of 
freedom, the energy scales separate and the orbital order occurs in 
KCuF$_{3}$ below the structural phase transition at rather high 
temperature, $T_{\rm OO}\simeq 800$ K. This demonstrates a strong 
Jahn-Teller coupling between the orbitals and lattice distortions 
\cite{vdB99}, which plays also an important role in LaMnO$_3$ 
\cite{Fei99} and cannot be ignored when the data for real compounds 
are explained. It has been also realized that the so-called Goodenough 
processes \cite{Jef92,Mos02,Ole05}, involving excitations on oxygen 
(ligand) sites, do contribute to the superexchange in charge-transfer 
insulators, and the structure of the effective Hamiltonian is richer 
than that of the KK model. However, we shall consider here just the 
spin-orbital superexchange models as they arise in the original 
derivation \cite{Ole00} from the multiband Hubbard model \cite{Ole83}.

While the coexisting $A$-AF order and the $C$-type orbital order 
($C$-OO) are well established in KCuF$_{3}$ below the N\'eel temperature
$T_{N}\simeq 39$ K and this phase is reproduced by the spin-orbital 
$d^9$ superexchange model in the MF approximation \cite{Ole00}, the 
phase diagram of this model beyond the MF approach is still unknown 
because of strongly coupled spin and orbital degrees of freedom, and 
poses an outstanding question in the theory: 
Which types of coexisting spin and orbital order (or disorder) are 
possible here when the microscopic parameters: 
($i$) crystal-field (CF) splitting of the $e_g$ orbitals $E_z$, and 
($i$) Hund's exchange $J_H$, are varied? 
It has been suggested that the long-range AF order is destroyed by 
spin-orbital quantum fluctuations \cite{Fei97}, but this is still 
controvesial and also other ordered states stabilized by the 
order-out-of-disorder mechanism might arise \cite{Kha97}. 

The purpose of this paper is to investigate the phase diagrams of the 
KK model at increasing dimensionality. Thereby we summarize the 
results of the earlier studies for a 2D monolayer \cite{Brz12}, 
bilayer \cite{Brz11}, and a 3D perovskite (cubic) system \cite{Brz13}. 
We show below that spin-orbital fluctuations and entanglement  
\cite{Ole12,Ole06} plays a very important role here and stabilizes 
exotic types of magnetic order in all these systems. We start with
introducing the KK model in Sec. \ref{sec:model}. Next in Sec. 
\ref{sec:mf} we explain two standard methods used to investigate the 
phase stability in different parameter regimes:
$(i)$ a one-site MF approximation (Sec. \ref{sec:ssmf}), and 
$(ii)$ a cluster MF approximation (Sec. \ref{sec:cmf}). 
We show the essential differences between the phase diagrams obtained 
in these two methods and argue that new types of exotic spin order 
arise from the entangled spin-orbital interactions. We also address the 
types of order found at finite temperature in the 2D monolayer, where 
we show that the magnetic order is more fragile even in the absence of 
the Jahn-Teller coupling. In Sec. \ref{sec:enta} we explain the origin 
of the spin order found in the cluster MF approximation. The paper is 
summarized in Sec. \ref{sec:summa}.

\section{Kugel-Khomskii model}
\label{sec:model}

The spin-orbital superexchange KK model was originally introduced for 
Cu$^{2+}$ ($d^9$) ions in the perovskite structure of KCuF$_3$, with 
$S=1/2$ spins and $e_g$ orbital pseudospins $\tau=1/2$ pseudospins
\cite{Kug73}. The correct multiplet structure was included only later
\cite{Ole00} when it was derived from the degenerate Hubbard Hamiltonian 
with hopping $t$, intraorbital Coulomb interaction $U$ and Hund's 
exchange $J_H$ for $e_g$ electrons \cite{Ole83}. It describes the 
Heisenberg SU(2) spin interactions coupled to the orbital problem, 
with the superexchange constant $J=4t^2/U$,
\begin{eqnarray} 
\label{kk}
{\cal H}&=&-\frac{1}{2}J\!\!\sum_{\langle ij\rangle||\gamma}
\left\{\left(r_1\,\Pi^t_{\langle ij\rangle}+r_2\,\Pi^s_{\langle
ij\rangle}\right)
\left(\frac{1}{4}-\tau^{\gamma}_i\tau^{\gamma}_j\right)\right. \nonumber \\
&+&\left. (r_2+r_4)\,\Pi^s_{\langle ij\rangle}
\left(\frac{1}{2}-\tau^{\gamma}_i\right)
\left(\frac{1}{2}-\tau^{\gamma}_j\right)\right\}+ {\cal H}_0\,.
\end{eqnarray}
where the CF splitting term,
\begin{equation} 
\label{H0}
{\cal H}_0=E_z\sum_{i}\tau_i^z,
\end{equation}
lifts the orbital degeneracy, and the coefficients, 
\begin{equation} 
r_{1}=\frac{1}{1-3\eta},\hskip.5cm
r_{2}=\frac{1}{1- \eta},\hskip.5cm
r_{4}=\frac{1}{1+\eta},
\label{allr}
\end{equation}
($r_3=r_2$) depend on Hund's exchange \cite{Ole00},
\begin{equation} 
\label{eta}
\eta\equiv\frac{J_H}{U}.
\end{equation}
Here $\gamma=a,b,c$ is the bond direction along one of the cubic axes 
$\gamma$. In a bilayer two $ab$ planes are connected by interlayer bonds 
along the $c$ axis \cite{Brz11}, while a monolayer has only bonds within 
a single $ab$ plane \cite{Brz12}, i.e., $\gamma=a,b$. In all these cases 
spin interactions are described with the help of spin projection 
operators on a triplet or a singlet configuration on a bond 
$\langle ij\rangle$,   
\begin{eqnarray} 
\label{projects} 
\Pi_{\langle ij\rangle}^{s}=\frac{1}{4}-{\bf S}_i\cdot{\bf S}_{i+\gamma}, 
\\
\label{projectt} 
\Pi_{\langle ij\rangle}^{t}=\frac{3}{4}+{\bf S}_i\cdot{\bf S}_{i+\gamma},
\end{eqnarray}
and $\tau^{\gamma}_i$ are the orbital operators for bond direction 
$\gamma=a,b,c$. They are defined in terms of Pauli matrices in the 
orbital space, $\{\sigma^x_i,\sigma^z_i\}$, as follows:
\begin{eqnarray}
\tau^{a(b)}_i&\equiv&\frac{1}{4}\left(-\sigma^z_i\pm\sqrt{3}\sigma^x_i\right),
\\
\tau^c_i&\equiv&\frac{1}{2}\,\sigma^z_i=\frac{1}{2}\left(n_{iz}-n_{ix}\right).
\end{eqnarray}
These operators act in the orbital space, with the basis states:
\begin{equation}
\label{egbasis}
|z\rangle\equiv (3z^2-r^2)/\sqrt{6}, \hskip .7cm
|x\rangle\equiv ( x^2-y^2)/\sqrt{2},
\end{equation}
and the $|z\rangle$ ($|x\rangle$) orbital is an "up" ("down") orbital 
state. Finally, $E_z$ in Eq. (\ref{kk}) is the crystal-field (CF) 
splitting of $e_g$ orbitals which favors either $|z\rangle$ (if $E_z<0$) 
or $|x\rangle$ (if $E_z>0$) orbital state occupied by a hole at each site 
$i$. Thus the model Eq. (\ref{kk}) depends on two parameters: $E_z/J$ 
and Hund's exchange $\eta$ (\ref{eta}) --- we vary them below to 
determine the phase diagrams analyzed in Sec. \ref{sec:mf}.

\section{Mean-field phase diagrams}
\label{sec:mf}

\subsection{Spin and orbital order}
\label{sec:so}

\begin{figure}[t!]
\begin{minipage}[t]{0.49\columnwidth}%
\begin{center}\includegraphics[clip,width=4.1cm]{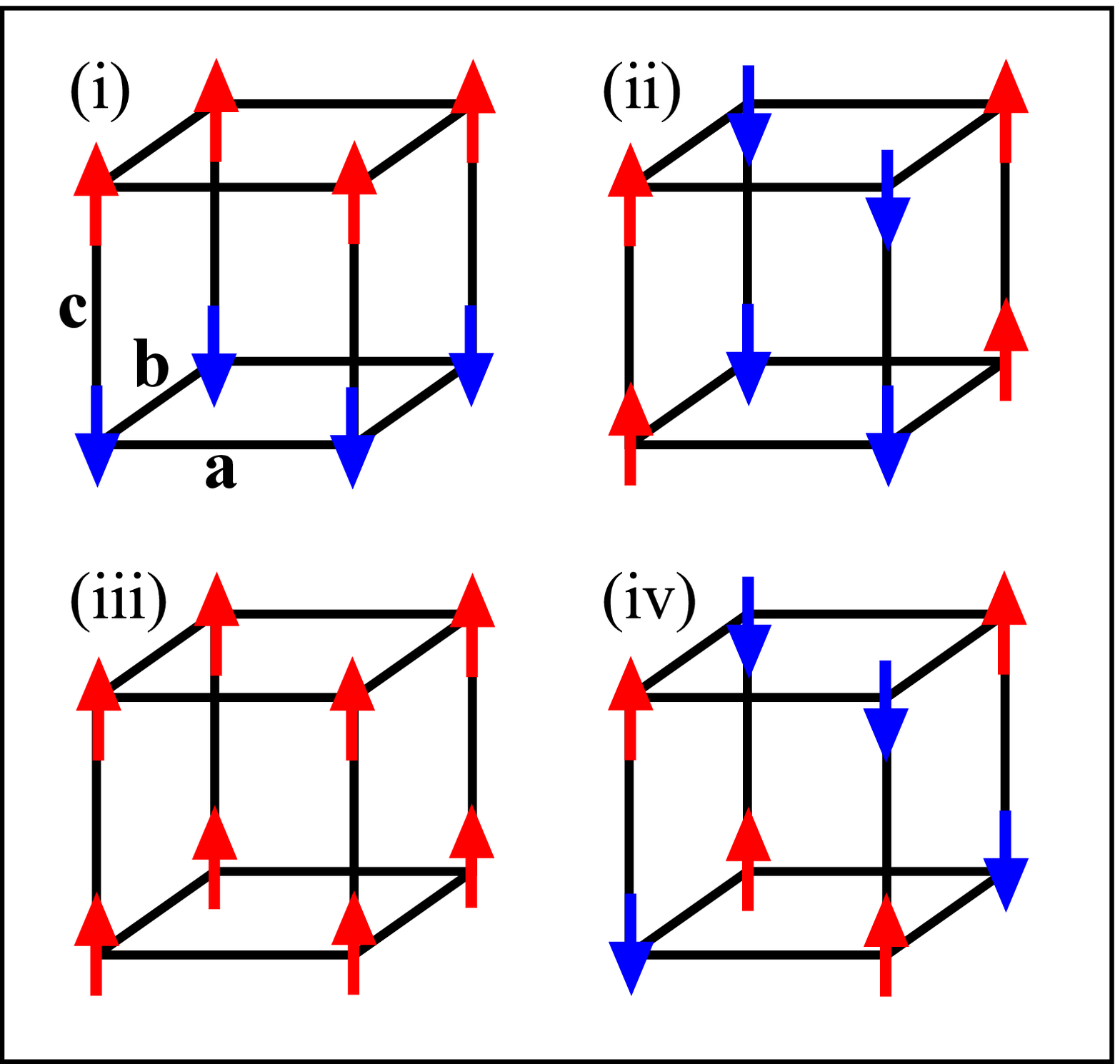}\end{center} 
\end{minipage}
\begin{minipage}[t]{0.49\columnwidth}%
\begin{center}\includegraphics[clip,width=3.8cm]{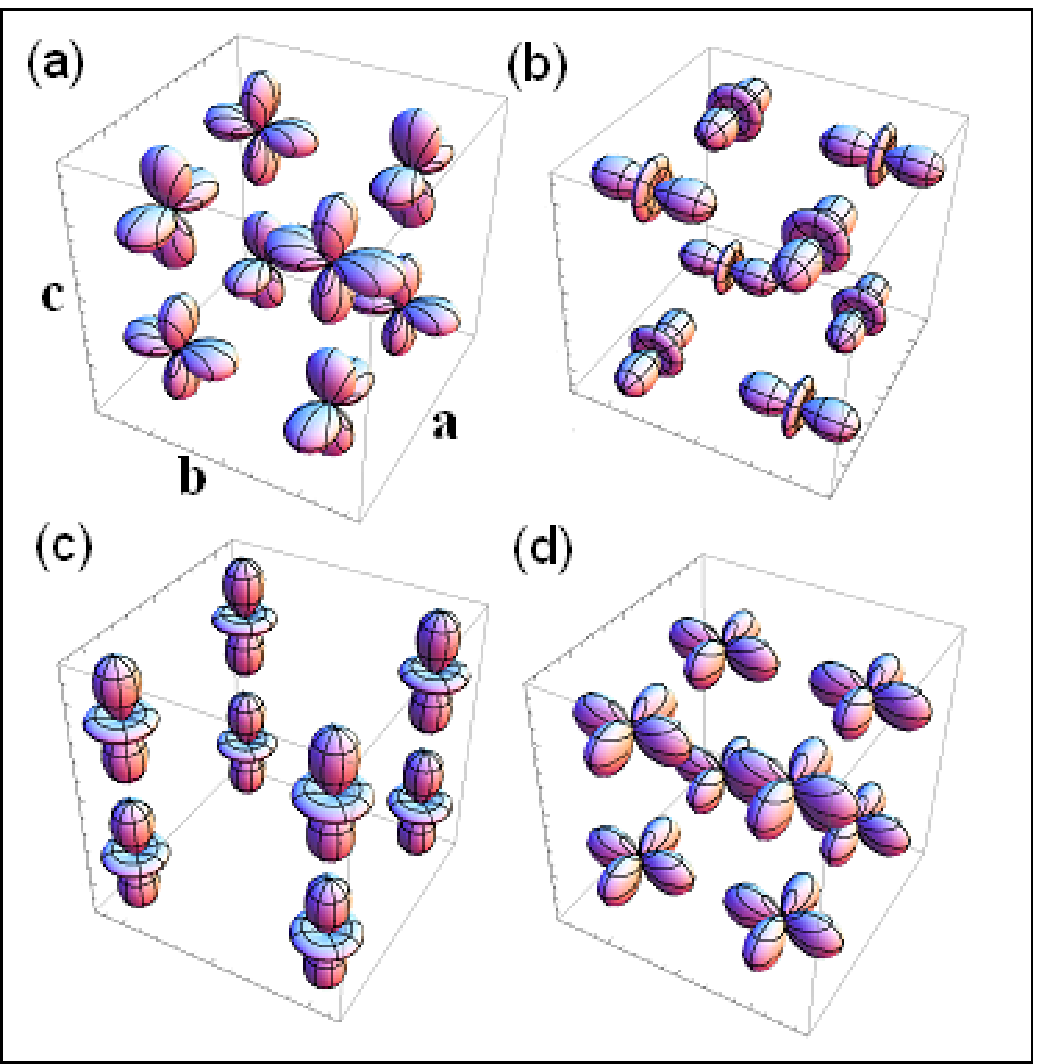}\end{center}
\end{minipage} 
\caption{Possible types of one- and two-sublattice spin and orbital 
order in a cubic system.
Left panel ---  
schematic view of four representative spin phases
(arrows stand for up or down spins):
(i)  $A$-AF,
(ii) $C$-AF,
(iii) FM, and
(iv) $G$-AF one.
Right panel --- 
schematic view of four types of orbital order
on a cube of the 3D (bilayer) lattice:
(a) AO order with $\langle\tau_{i}^{a(b)}\rangle=1/2$ changing from 
site to site and $\langle\tau_i^c\rangle=-1/4$, obtained for $E_{z}<0$,
(b) AO order with $\langle\tau_{i}^{a(b)}\rangle=-1/2$ changing from
site to site and $\langle\tau_i^c\rangle=1/4$, obtained for $E_{z}>0$,
(c) FO order with occupied $z$ orbitals and
$\langle\tau_{i}^{c}\rangle=1/2$, and
(d) FO order with occupied $x$ orbitals and
$\langle\tau_{i}^{c}\rangle=-1/2$.
}
\label{fig:4cubes}
\end{figure}

We consider spin and orbital orbital order with up to two sublattices,
as well as phases composed of equivalent cubes (or plaquettes in the 2D 
system). The spin order is simpler due to the SU(2) symmetry of the 
Heisenberg interactions, as it gives $\langle S_j^z\rangle=\pm 1/2$ when 
the $z$-th spin components are chosen to determine the broken symmetry 
state, so it suffices to consider AF or FM bonds in 
various nonequivalent phases with long-range spin order. 
The spin SU(2) interaction is replaced in the MF approximation by 
local operators at site $i$ interacting with the MF values for the 
operators at neighboring sites $j$,
\begin{equation}
\label{mfs}
{\vec S}_i{\vec S}_j \simeq S_i^z\langle S_j^z\rangle 
+\langle S_i^z\rangle S_j^z -\langle S_i^z\rangle\langle S_j^z\rangle. 
\end{equation}
We consider the magnetic phases depicted in a schematic way in Fig. 
\ref{fig:4cubes}, where $\langle S_j^z\rangle\neq 0$:
($i$) $A$-AF phase --- with FM order in the $ab$ planes and AF
correlations along the $c$ axis [Fig. \ref{fig:4cubes}(i)],
($ii$) $C$-AF  phase --- with AF order in the $ab$ planes and FM
correlations along the $c$ axis [Fig. \ref{fig:4cubes}(ii)],
($iii$) FM phase [Fig. \ref{fig:4cubes}(iii)], and
($iv$) $G$-AF phase N\'eel state [Fig. \ref{fig:4cubes}(iv)].
The first two phases, ($i$) $A$-AF and ($ii$) $C$-AF phase, contain 
partly FM bonds which are stabilized by the orbital order (OO).

On the contrary, due to the absence of SU(2) symmetry, the OO in the 
present KK models may involve not only one of the two basis $e_g$ 
orbital states $\{|z\rangle,x\rangle\}$, but also their linear 
combinations,
\begin{equation}
\label{mixing}
|\theta\rangle=\cos\left(\frac{\theta}{2}\right)|z\rangle
              +\sin\left(\frac{\theta}{2}\right)|x\rangle,
\end{equation}
parametrized by an angle $\theta$ which defines their amplitudes at 
site $i$. In the orbital sector we apply then the MF decoupling for 
the products $\{\tau_i^{\gamma}\tau_{i\pm\gamma}^{\gamma}\}$ along 
the axis $\gamma$:
\begin{equation}
\label{mft}
\tau_i^{\gamma}\tau_{i\pm\gamma}^{\gamma}
\simeq \langle \tau_i^{\gamma} \rangle \tau_{i\pm\gamma}^{\gamma}+
\tau_i^{\gamma}\langle \tau_{i\pm\gamma}^{\gamma}\rangle -\langle 
\tau_i^{\gamma} \rangle\langle \tau_{i\pm\gamma}^{\gamma}\rangle.
\end{equation}
As order parameters we take $t^a\equiv\langle\tau_1^{a}\rangle$
and $t^c\equiv\langle\tau_1^{c}\rangle$ for a chosen site $i=1$
(which is sufficient in orbital sector as $t^b=-t^a-t^c$) and we
assume two orbital sublattices: each neighbor of the site $i$ is
rotated by $\pi/2$ in the $ab$ plane meaning that
$\langle\tau_{i+\gamma}^{a(b)}\rangle=t^{b(a)}$. 
A frequently encountered form of the OO is a 
two-sublattice structure, with two orbitals given by angles 
$\theta_A=\theta$ and $\theta_B=-\theta$:
\begin{eqnarray}
\label{aaf}
|\theta_A\rangle_i&=&\cos\left(\frac{\theta}{2}\right)|z\rangle_i
                    +\sin\left(\frac{\theta}{2}\right)|x\rangle_i,
\nonumber\\
|\theta_B\rangle_j&=&\cos\left(\frac{\theta}{2}\right)|z\rangle_j
                    -\sin\left(\frac{\theta}{2}\right)|x\rangle_j.
\end{eqnarray}
Examples of simple OO which may {\it a priori\/} coexist with spin 
order are shown as well in Fig. \ref{fig:4cubes}:
alternating orbital (AO) order with either $x$-like or $z$-like
orbitals in (a) and (b), and two ferro-orbital (FO) orders with either 
$z$ or $x$ orbitals occupied at each site, in (c) and (d). In reality 
the angles for the AO states vary in a continuous way as functions of 
the CF parameter $E_z/J$, and therefore an efficient way of solving 
the one-site MF equations consists of assuming possible magnetic 
orders and for each of them deriving the effective orbital-only 
model. Such models are next compared and the phases with the lowest 
energy is found, as presented below for the case of the 2D monolayer, 
\cite{Brz12}, the KK bilayer \cite{Brz11}, and for the 3D cubic model 
\cite{Brz13}. 

The simplest approximation to obtain the possible types of order in a 
spin-orbital model, like the KK models Eq. (\ref{kk}) considered here, 
is the single-site MF approach which consists of two steps:
$(i)$ decoupling of spin and orbital interactions, and 
$(ii)$ subsequent factorization of interactions of both types on the 
bonds into (spin or orbital) operators at a given site $i$ coupled to 
the MF terms on its neighboring sites.  
The values of the projection operators (\ref{projects}) and 
(\ref{projectt}) depend on the assumed spin order, and they 
may be easily eliminated in this approach when the spin scalar 
products are replaced by their values in the MF states, being either 
$\langle {\bf S}_i\cdot{\bf S}_{i+\gamma}\rangle_{\rm FM}= 1/4$, or  
$\langle {\bf S}_i\cdot{\bf S}_{i+\gamma}\rangle_{\rm AF}=-1/4$. Taking 
different types of spin order ($i$)-($iv$), and assuming the classical 
average values of the spin projection operators, one finds the MF
equations which are next solved for each of the considered three 
systems: the 2D monolayer, the bilayer, and the 3D perovskite. 
Solutions of the self-consistency equations and ground state energies 
in different phases can be obtained analytically, as explained on the 
example of the bilayer system in \cite{Brz11}.

\subsection{Single-site mean-field approximation}
\label{sec:ssmf}

The simplest approach is a single-site MF approximation applied to the 
KK models Eq. (\ref{kk}). It excludes any spin fluctuations as the spin 
projectors $\Pi^{t(s)}_{\langle ij\rangle}$ 
($\Pi^s_{\langle ij\rangle}$) are here replaced by their mean values, 
where the dependence on the bond $\langle ij\rangle$ reduces to 
direction $\gamma$ in phases with translationally invariant magnetic 
order shown in Fig. \ref{fig:4cubes}. The phase diagrams of the KK 
model follow from the two competing trends when the parameters, the 
CF splitting $E_z/J$, and Hund's exchange  $\eta$ (\ref{eta}) are 
varied. While increasing CF parameter $E_z$ causes a switch from $z$ to 
$x$ orbitals, increasing Hund's exchange favors FM interactions along 
at least the $c$ axis, and coexisting with the AO phase.

\begin{figure}[t!]
\includegraphics[width=8.0cm]{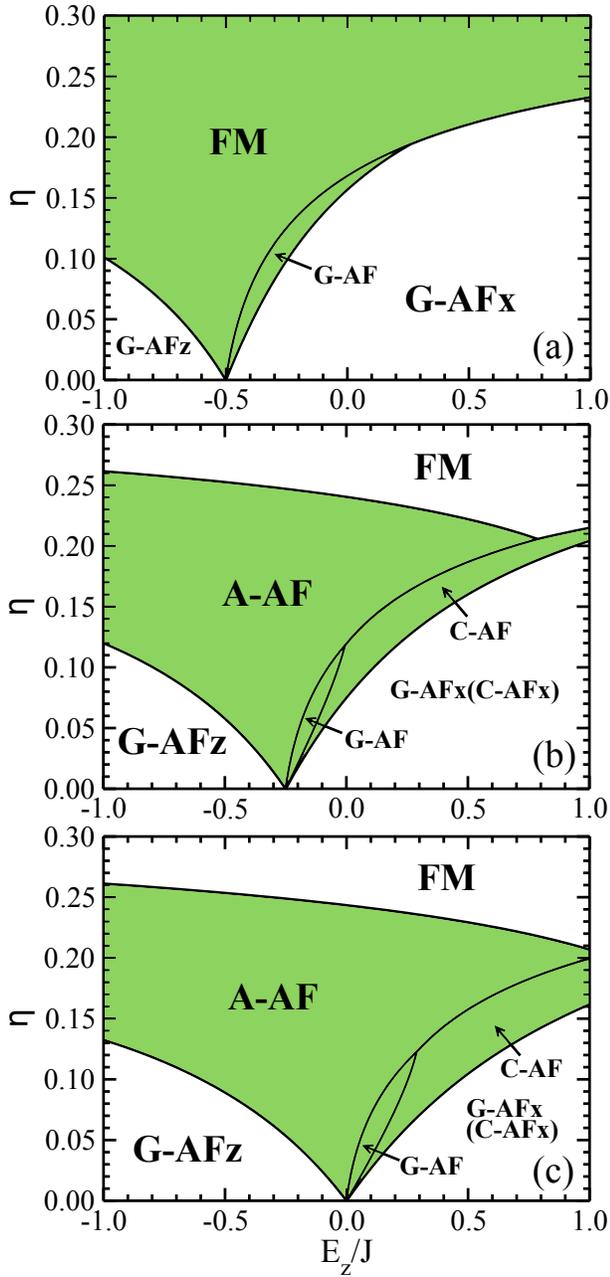} 
\caption{
Phase diagrams of the KK models in the single-site MF approximation as
obtained for:
(a) the 2D monolayer \cite{Brz12},
(b) the bilayer \cite{Brz11}, and
(c) the 3D perovskite system \cite{Brz13}. 
Shaded dark gray (green) areas indicate phases with AO order:
FM and $G$-AF phase for the monolayer (a), and the $A$-AF, $C$-AF and 
$G$-AF phases for the bilayer (b) and the 3D perovskite (c) [here the 
FM phases which coexist also with the AO order are not shadded for 
clarity]. The remaining $G$-AF phases ($G$-AF$x$ and $G$-AF$z$) are 
accompanied by FO order with fully polarized orbitals, either $x$ 
(for $E_z>E_z^0$) or $z$ (for $E_z<E_z^0$). The quantum critical
point with degenerate $G$-AF$x$, $G$-AF, $A$-AF, and $G$-AF$z$ phases 
is found at $(E_z,\eta)=(E_z^0,0)$, with decreasing $E_z^0$ from the 
3D perovskite via the bilayer to the 2D monolayer.
}
\label{fig:ssmf}
\end{figure}

\begin{figure}[t!]
\includegraphics[width=8.0cm]{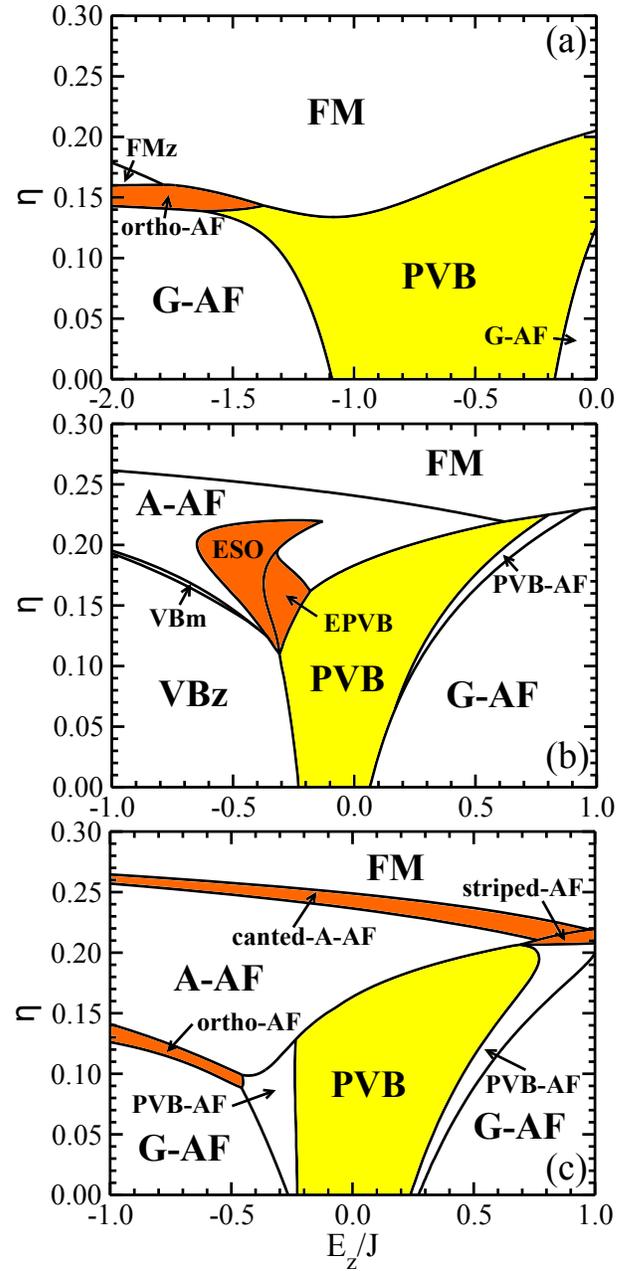}
\caption{
Phase diagrams of the KK models in the cluster MF approximation as
obtained for:
(a) the 2D monolayer \cite{Brz12},
(b) the bilayer \cite{Brz11}, and
(c) the 3D perovskite system \cite{Brz13}. 
Plaquette valence-bond (PVB) phase with alternating spin singlets in
the $ab$ planes is highlighted in light gray (yellow) --- it occurs 
between the phases with long-range magnetic order, and the QCP is 
hidden. When strong competition between AF and FM interactions occurs 
at increasing $\eta$, new phases with exotic spin order are found --- 
they are highlighted in dark gray (orange).
}
\label{fig:cmf}
\end{figure}

Using the factorized form of the 
spin-orbital superexchange, and the order parameters,
\begin{equation}
s_{i}^{z}\equiv\left\langle S_{i}^{z}\right\rangle ,
\hskip .7cm
t_{i}^{\gamma}\equiv\left\langle \tau_{i}^{\gamma}\right\rangle ,
\label{ops1}
\end{equation}
with $|\langle S_j^z\rangle|=\frac12$, we determined MF energies of all 
possible phases with spin long-range order and optimized values of the 
orbital order parameters $\{t_i^{\gamma}\}$.
Consider first the case of vanishing Hund's exchange $\eta=0$. In this 
case the multiplet structure (\ref{allr}) collapses to a single level
and $r_i=1$ $\forall i$. Therefore the excited states with double
occupancies in two different orbitals, e.g. $x_i^1z_i^1$ at site $i$, 
do not introduce any spin dependence of the superexchange as the 
contributions from the triplet and singlet excitation compensate each 
other, and the only magnetic term stems from double occupancies at the 
same $e_g$ orbital. Such terms are AF and resemble the ones derived 
from the Hubbard model without orbital degeneracy in the $t$-$J$ model 
\cite{Cha77}. Therefore, one finds that at $\eta=0$ two AF N\'eel 
phases are degenerate: $G$-AF$z$ and $G$-AF$x$. Actually, individual 
contributions to the ground state energy from the bonds along the $ab$ 
axes and along the $c$ axis are quite different in both phases: 
while the $G$-AF$x$ phase is 2D, with no coupling between the $ab$ 
planes and is realized for instance in La$_2$CuO$_4$ \cite{Ole12}, the 
AF superexchange along the $c$ axis is stronger by a factor of 16 than 
the one along the bonds in the $a$ planes in the $G$-AF$z$ phase. 
Nevertheless, these energy contributions add to the same value in a 3D
cubic system \cite{Ole00} and the actual occupied orbital (FO order) is 
decided by the value of the CF splitting. One finds the $G$-AF$z$ phase 
for $E_z<0$ and $G$-AF$x$ phase for $E_z>0$ --- they are degenerate at 
the quantum critical point [QCP $(E_z^0,\eta)=(0,0)$]. This reflects the 
cubic symmetry of the model (\ref{kk}) at $E_z=0$.

In two other cases, the cubic symmetry is broken and the transition 
between the $G$-AF$z$ and $G$-AF$x$ phase (the QCP) occurs now at 
finite $E_z$: $E_z^0=-0.25J$ and $E_z^0=-0.5J$ for the bilayer [Fig. 
\ref{fig:ssmf}(b)] and for the 2D monolayer [Fig. \ref{fig:ssmf}(a)]. 
This follows from the anisotropic superexchange in the $G$-AF phases
--- the magnetic MF energy decreases rapidly in the $G$-AF$z$ phase 
with weak exchange bonds and has to be compensated by the CF term. 

Having two degenerate phases at $(E_z^0,0)$ in any of the considered 
cases, it suffices to add an infinitesimal Hund's exchange $\eta>0$ to 
destabilize the N\'eel AF order in favor of the $A$-AF phase, with FM 
interactions in the $ab$ planes and AF ones along the $c$ axis. 
Such anisotropic magnetic interactions are supported by the AO order,
with two different orbitals along each bond $(ij)\in ab$, as shown in 
Fig. \ref{fig:4cubes}(a). The occupied orbitals belong to two 
sublattices $A$ and $B$, as given in Eqs. (\ref{aaf}). One may wonder 
whether the $G$-AO order depicted in Fig. \ref{fig:4cubes}(a) coexists 
indeed with the $A$-AF order, and the answer to this question goes 
beyond the superexchange model (\ref{kk}). In fact, a small interaction 
with the lattice (ignored in the present analysis) selects one of these 
phases as the orbital alternation implies lattice distortions around 
each Cu$^{2+}$ ion. Such distortions occur within the $ab$ 
planes and a better energy is found when the planes are repeated along  
the $c$ axis --- this implies the $C$-AO order coexisting with 
the $A$-AF phase. 

As expected, the spin order is $G$-AF when the CF term is large and 
favors one of the two orbitals, while FM interactions appear when the 
OO changes to AOs and the spin interactions in the $ab$ planes become 
FM. In contrast to the FO phases, the AO order in the shaded phases in 
Fig. \ref{fig:ssmf} is more involved and the angle $\theta$ in Eqs. 
(\ref{aaf}) is selected by the energy minimization in the respective 
phase. In most cases the transitions between the $G$-AF and $A$-AF
(or $C$-AF or $FM$) order are first order.
The $A$-AF phase, stable here in a broad range of parameters both in 
the bilayer [Fig. \ref{fig:ssmf}(b)] and in the 3D perovskite [Fig. 
\ref{fig:ssmf}(c)], develops from the FM phase in the case of the 2D 
monolayer [\ref{fig:ssmf}(a)]. In the two former systems another 
transition to the (anisotropic) FM phase occurs at large 
$\eta\simeq 0.25$, when the AF terms in the superexchange terms are
dominated by triplet charge excitations. In all the cases the $G$-AF 
sp[in order occurs as a precursor of the $A$-AF (FM) phase, when $E_z$ 
is decreased and the $ab$ planes become weakly coupled by the AO order.
The $C$-AF phase occurs in addition in a narrow range of parameters in 
between the $G$-AF$x$ and the $A$-AF phase in Figs.
\ref{fig:ssmf}(b) and \ref{fig:ssmf}(c). This phase is unexpected and 
suggests that the phase diagrams derived in better approximations 
might be quite different.

\subsection{Cluster mean-field approximation}
\label{sec:cmf}

Knowing that spin-orbital quantum fluctuations are enhanced near 
orbital degeneracy \cite{Fei97}, it is of crucial importance to include 
them when the phase diagrams of the KK models are investigated. Exact 
diagonalization gives exact ground states of finite clusters and may 
be combined with MF approach when a cluster under consideration is in
contact with its neighboring clusters via the MF terms. We developed 
this so-called cluster MF approximation for the KK models by embedding 
a four-site cluster in the plaquette MF (PMF) used in the 2D model 
\cite{Brz12}, and a cube in the bilayer case \cite{Brz11}. The most 
natural choice for the 3D perovskite is a cube as well, but here we 
limit ourselves to four-site clusters \cite{Brz13} 
(a plaquette or a linear cluster) to avoid tedious numerical analysis.  

The interactions along bonds which belong to a cluster considered in 
each case are treated by exact diagonalization, while the bonds which 
couple the cluster with its neighbors are decoupled in the MF 
approximation. In this way we arrive at the self-consistent MF 
equations for the order parameters:
\begin{equation}
s_{i}^{\alpha}\equiv\left\langle S_{i}^{\alpha}\right\rangle ,
\hskip .5cm
t_{i}^{\gamma}\equiv\left\langle \tau_{i}^{\gamma}\right\rangle ,
\hskip .5cm
v_{i}^{\alpha,\gamma}\equiv
\left\langle S_{i}^{\alpha}\tau_{i}^{\gamma}\right\rangle\,.
\label{ops2}
\end{equation}
Here we consider two spin components, $\alpha=x,z$, as the SU$(2)$ 
symmetry of the spin interactions may be broken in a more general way 
to include some exotic types of magnetic order obtained in particular
for the 2D monolayer \cite{Brz12}.
The mixed order parameters $\{v_{i}^{\alpha,\gamma}\}$ are essential 
here and influence the stability of phases by including on-site 
spin-orbital entanglement \cite{Brz11}. 

The essential qualitative difference between the cluster MF (Fig. 
\ref{fig:cmf}) and the single-site MF approach (Fig. \ref{fig:ssmf}) 
is the possibility of spin disorder, realized
in between the phases with spontaneously broken symmetry. 
{\it De facto\/}, the phases with long-range spin order ($G$-AF and 
$A$-AF ones) which coexist at the QCPs are replaced by the plaquette 
valence-bond (PVB) states, as shown in Fig. \ref{fig:cmf}. These states 
are characterized by local order on individual plaquettes within the 
$ab$ planes, with spin singlets coexisting with pairs of directional 
orbitals, $3x^2-r^2$ or $3y^2-r^2$, along the same bonds, and rather 
weak coupling between them. The orbital states behave more classically 
and fluctuations between different orientations of singlets are blocked 
on each plaquette by the MF terms which couple this
plaquette to its neighbors. By considering different possible covering 
of the lattice by such PVB states we could establish their alternation 
within the $ab$ planes, with each pair of neighboring plaquettes 
forming a superlattice of plaquettes with alternating horizontal and
vertical singlet bonds. These states are particularly robust in the 2D
monolayer and suppress the FM order in a broad regime of parameters, 
as they are not disturbed here by the perpendicular bonds along the $c$ 
axis, present in the bilayer and in the 3D perovskite. 

The QCPs in the single-site MF approximation and the superlattice of 
alternating plaquettes realized in the cluster MF may be seen as 
indications of frustrated spin-orbital interactions. It is surprising 
that this frustration leads not only to spin disorder but also to 
rather exotic types of spin order, in particular in the 2D monolayer 
and in the 3D perovskite lattice. In both cases the ortho-AF phase is 
found in between the $G$-AF and $A$-AF phases, and in the 3D model in
addition also the striped-AF and canted-$A$-AF phases are stable in the 
vicinity of the FM order. Although the present KK models contain only 
nearest-neighbor (NN) superexchange interactions, when the AF and FM 
contributions compete, the usual NN superexchange becomes ineffective 
and other higher order processes contribute \cite{Brz12,Brz13}, as 
explained below in Sec. \ref{sec:enta}. The case of the bilayer is 
different as the bonds along the $c$ axis contribute here with the full 
spin singlet energy when the CF term ($E_z<0$) selects the $z$ orbitals. 
This dominates over the other terms along the $ab$ bonds and favors 
phases with spin disorder in this regime of parameters: entangled 
spin-orbital phase (ESO) and entangled PVB (EPVB) phase \cite{Brz11}, 
shown in Fig. \ref{fig:cmf}(b).

\subsection{Phase diagrams for the monolayer at $T>0$}
\label{sec:mono}

\begin{figure}[t!]
    \includegraphics[width=8.0 cm]{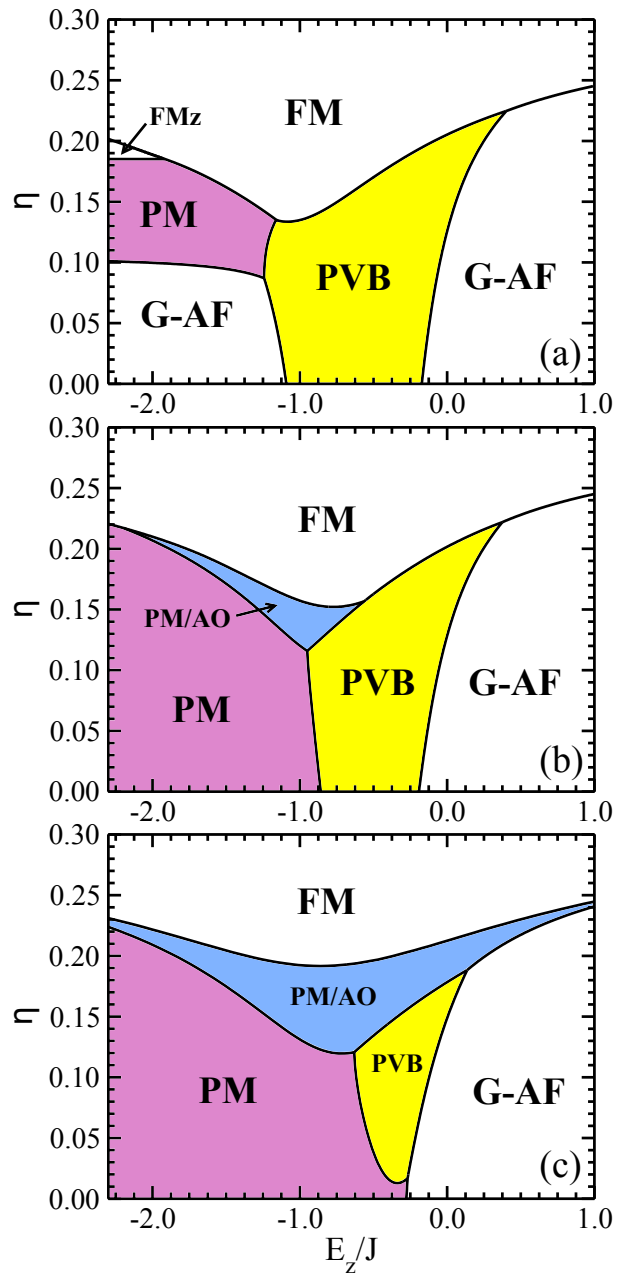}
\caption{
Evolution of the phase diagram of the 2D KK model with increasing 
temperature:
(a) $T=0.05J$, 
(b) $T=0.20J$, and 
(c) $T=0.34J$. 
The spin order changes gradually to paramagnetic (PM) under increasing 
temperature while the OO is more robust.
}
\label{fig:T>0}
\end{figure}

Before analyzing spin-orbital entangled states at temperature $T=0$, we 
emphasize that the spin and orbital interactions concern quite different 
energy scales. As an illustrative example we consider the 2D monolayer, 
where the phase diagrams obtained at finite $T$ demonstrate that the spin 
order is robust only in two cases: 
($i$) the $G$-AF order due to large superexchange at $E_z>0$, or 
($ii$) the FM order stabilized by large Hund's exchange $\eta>0.2$. The 
other spin ordered phases are stronger influenced by thermodynamic spin
fluctuations because the exchange interactions are much weaker in them. 
The exotic ortho-AF phase disappears already at $T=0.05J$ --- the  
paramagnetic (PM) spins are then accompanied by the FO order of the $z$ 
states, selected by the CF term $E_z<0$, see Fig. \ref{fig:T>0}(a). In
contrast, both phases with spin long-range $G$-AF and FM$z$ order which 
are away from the AF$\leftrightarrow$FM transition in the superexchange, 
are more stable. 

Due to the orbital shape, the exchange interactions between pairs of 
$z$ orbitals are much weaker, and both phases with this type of FO 
order, the $G$-AF$z$ and FM$z$ phase, are not found at $T=0.20J$, see 
Fig. \ref{fig:T>0}(b). Here a new PM 
phase with AO order (PM/AO phase) appears in cases when Hund's exchange 
is not strong enough to give robust FM order. At the same time the PVB 
phase shrinks as the spins are too weakly coupled to form stable spin 
singlets. Even larger regimes of the PM and PM/AO phase are found when 
temperature is further increased to $T=0.34J$, see Fig. \ref{fig:T>0}(c).

\section{Spin-orbital entangled states}
\label{sec:enta}

\subsection{Origin of exotic magnetic orders}
\label{sec:origin}

We explain the origin of the exotic magnetic order on the example of the 
ortho-AF phase in the 2D monolayer. This phase occurs in between the 
$G$-AF$z$ phase and the FM phase (either FM$z$ or with the AO order), 
see Fig. \ref{fig:cmf}(a). Large negative CF splitting, $E_z<0$, favors 
there $z$ orbitals, and the CF term (\ref{H0} may be treated as the the 
unperturbed part of the Hamiltonian, while the superexchange $\propto J$ 
is the perturbation, ${\cal V}\equiv {\cal H}-{\cal H}_{0}$. 
The ground state $|0\rangle$ of ${\cal H}_0$ is the FO$z$ state with $z$ 
orbitals occupied by a hole at each site,  
$\tau_{i}^{c}\left|0\right\rangle=\frac{1}{2}\left|0\right\rangle$, and 
the spin order is undetermined. Orbital excitations have a large gap
and the ratio $J/|E_z|\equiv|\varepsilon_z|^{-1}$ is a small parameter 
which may be used here to construct the expansion in powers of 
$|\varepsilon_z|^{-1}$,
\begin{equation}
\label{Hs}
H_{s}\simeq J\left\{H_s^{(1)}+H_s^{(2)}+H_s^{(3)}\right\},
\end{equation}
The terms $H_s^{(n)}$ are spin interaction in $n$-th order of this 
perturbative expansion. The first order term is an average
$H_{s}^{(1)}\equiv\langle 0|{\cal V}|0\rangle$ which is just the 
superexchange in Eq. (\ref{kk}), projected on the ground state 
$|0\rangle$, i.e., obtained for the FO$z$ state. As expected, this term
is the Heisenberg interaction along the bonds $\langle ij\rangle\in ab$,
\begin{equation}
\label{Hs1}
H_{s}^{(1)}=\frac{1}{2^{5}}\left(-3r_{1}+4r_{2}+r_{4}\right)
\sum_{\left\langle ij\right\rangle }
\left({\bf S}_{i}\cdot{\bf S}_{j}\right).
\end{equation}
For $\eta<0.155$ ($\eta>0.155$) the spin interaction is AF (FM), and as
long as no further contributions are evaluated the AF$\leftrightarrow$FM
transition takes place at $\eta_0=0.155$. 

Yet, the magnetic order is determined by higher order $H_s^{(2)}$ and 
$H_s^{(3)}$ terms for $\eta_0\simeq 0.155$. These interactions involve 
more than two sites and are obtained by considering all excited states
$|n\rangle$, with orbitals flipped from $z$ to $x$ --- they can be 
evaluated from the matrix elements involving excited states, 
$\langle n|{\cal V}|0\rangle$ \cite{Brz12}. We determined them 
taking certain number of $z$-orbitals flipped to $x$-orbitals 
and derived their average values. All the averages are taken between 
orbital states to derive the spin model Eq. (\ref{Hs}).
Given that ${\cal V}$ has non-zero overlap only with states having one 
or two NN orbitals flipped from $z$ to $x$, one finds in second order
effective interactions which couple next nearest neighbors (NNN) and 
third nearest neighbors (3NN) in the lattice \cite{Brz12}.

\begin{figure}[t!]
    \includegraphics[width=8.2 cm]{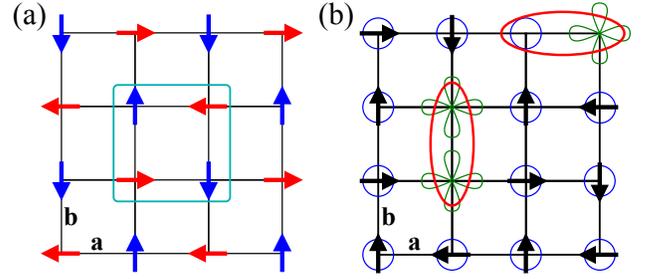}
\caption{
Artist's view of the exotic spin order found in the ortho-AF phase in 
the 2D monolayer:
(a) one of the two nonequivalent spin configurations, with four spin
sublattices --- up/down arrows stand for eigenstates of 
$\left\langle S_{i}^{z}\right\rangle =\pm\frac12$, 
while right/left arrows for 
$\left\langle S_{i}^{x}\right\rangle =\pm\frac12$;
(b) orbital FO$z$ order is locally modified by orbital excitations 
from $|z\rangle$ (circles) to $|x\rangle$ orbitals (clovers), and
the ortho-AF spin order (arrows) is then locally replaced by spin 
singlets (ovals).
} 
\label{fig:2D}
\end{figure}

The NNN interaction in $H_s^{(2)}$ are AF and would give two quantum 
antiferromagnets on interpenetrating sublattices. To explain the spin 
ortho-AF order shown in Fig. \ref{fig:2D}(a), i.e., NN spins being 
perpendicular, one has to include the third order $H_s^{(3)}$ as well.
Qualitatively new terms as compared to the lower orders arise then 
\cite{Brz12}, with connected products of three different Heisenberg 
bonds. They provide four-spin couplings and modify the ground state 
energy. The final result of such an analysis is that the classical 
energy is indeed minimized by the configurations with angles 
$\varphi=\pi/2$ between the NN spins, as shown in Fig. \ref{fig:2D}(a). 

It is a challenge to write down the ground state of $H_s$ (\ref{Hs}), 
$\left|\mathrm{AF}_{\perp}\right\rangle$, using the described 
perturbative scheme. It turns out that the quantum corrections obtained 
within the spin-wave expansion are small and the spin state is nearly
classical. Nevertheless, the spins in $H_s$ are always dressed with 
orbital and spin-orbital fluctuations, and the ground state is rather
complex. Indeed, within the perturbative treatment one obtains the 
full spin-orbital ground state shown in Fig. \ref{fig:2D}(b), 
\begin{equation}
\label{so}
\left|\Psi_{\rm SO}\right\rangle\propto 
\left(1-
\sum_{n\not=0}   \frac{{\cal V}_n}{\varepsilon_{n}}+
\sum_{n,m\not=0} \frac{{\cal V}_n{\cal V}_m}{\varepsilon_{n}\varepsilon_{m}}
-\dots\right)\left|\Phi_0\right\rangle,
\end{equation}
where ${\cal V}_n\equiv\left|n\right\rangle\left\langle n\right|{\cal V}$, 
$\varepsilon_n$ are excitation energies, and 
$\left|\Phi_0\right\rangle\equiv|\mathrm{AF}_{\perp}\rangle|0\rangle$
is the disentangled classical (N\'eel-like) state of Fig. \ref{fig:2D}(a).
This classical state is dressed with both orbital and spin-orbital 
fluctuations via the terms which stem from the operator sum in front of 
$\left|\Phi_0\right\rangle$ in Eq. (\ref{so}). A simpler form is obtained 
when the purely orbital fluctuations are neglected and density of 
spin-orbital defects is assumed to be small, one finds 
\begin{equation}
\label{eq:ldAp}
\left|\Psi_{\rm SO}\right\rangle\simeq\,
\exp\left( -\frac{1}{|\varepsilon_z|}
\sum_{\langle ij\rangle||\gamma}{\cal D}_{ij}^{\gamma}\right) 
\left|\Phi_0\right\rangle,
\end{equation}
where
\begin{equation}
\label{Dij}
{\cal D}_{ij}^{\gamma}=\left\{
-A\,\sigma_{i}^{x}\sigma_{j}^{x}
+B\left(\sigma_{i}^{x}+\sigma_{j}^{x}\right)s_{\gamma}\right\}\Pi_{ij}^{s}
\end{equation}
is the spin-orbital excitation operator on the bond $\langle ij\rangle$, 
with $A=3(r_{1}+r_{4})/2^6$ and $B=\sqrt{3}(r_{1}+2r_{2}+3r_{4})/2^5$. 
Both terms in Eq. (\ref{Dij}) project on a NN spin singlet, but the 
first one flips two NN $z$-orbitals while the second one generates 
only one flipped orbital. The density of entangled defects in Fig. 
\ref{fig:2D}(b) increases when $|\varepsilon_z|$ is decreased and 
the ortho-AF phase is gradually destabilized. 

\begin{figure}[t!]
\begin{center}\includegraphics[width=8.2cm]{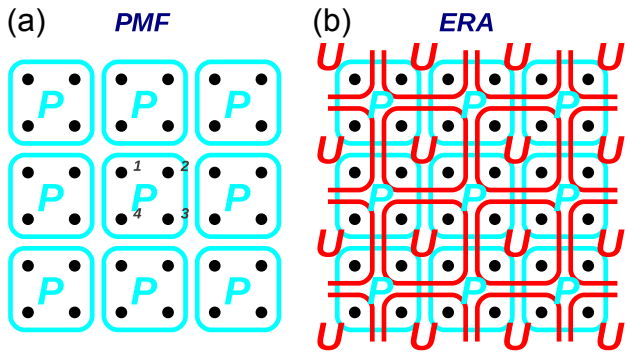}\end{center} 
\caption{
Two variational Ans\"atze used to study the ortho-AF phase: 
(a) the plaquette MF (PMF), and 
(b) the entanglement renormalization Ansatz (ERA). 
Black dots are lattice sites, ${\cal P}$'s are variational wave 
functions on $2\times2$ plaquettes, and ${\cal U}$'s are variational 
$2\times2$ unitary disentanglers. The difference between PMF and ERA 
is the additional layer of disentanglers between the plaquette wave 
functions ${\cal P}$ and the physical degrees of freedom. Their role 
is to introduce some entanglement between different plaquettes or to 
disentangle partially the plaquettes before the plaquette product 
Ansatz is applied. 
This figure is reproduced from \cite{Brz12}.
}
\label{fig:ERA} 
\end{figure}

Finally, it is worth to mention that the exotic ortho-AF order was not 
only predicted by the PMF and explained by the perturbative expansion, 
but its existence was also corroborated by a variational calculation 
with the Entanglement Renormalization Ansatz (ERA). It is remarkable 
that the range of stability of the ortho-AF phase is similar in both 
Ans\"atze in Fig. \ref{fig:ERA}. Actually, this exotic phase is even 
more robust in the ERA than in the PMF. 

\subsection{Examples of 3D exotic spin order}
\label{sec:ex}

In the 3D perovskite lattice the ortho-AF phase is found as well in the 
narrow range of $(E_z,\eta)$ where the spin order changes classically 
from the $G$-AF to $A$-AF one, and the order is FO$z$ \ref{fig:2D}(a) 
in $ab$ planes, and AF between the consecutive planes along the $c$ 
axis. 

\begin{figure}[t!]
    \includegraphics[width=8.2cm]{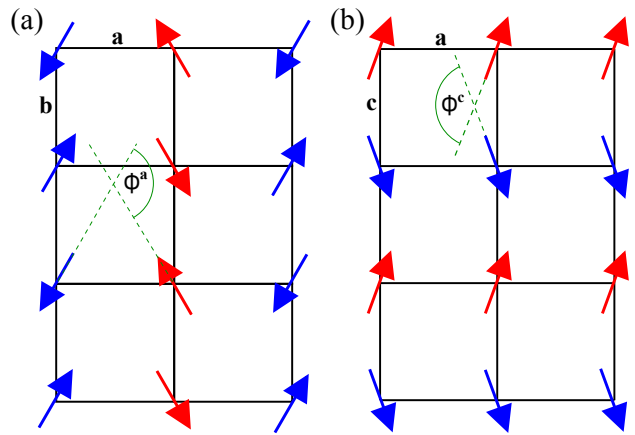}
\caption{
Two exotic spin orders realized by the 3D perovskite KK model at large 
Hund's exchange $\eta>0.2$:
(a) striped-AF order, with AF order along the $c$ axis and angle 
$\phi^a$ between the NN spins along the $a$ axis (spins are AF along 
the $b$ axis);
(b) canted-$A$-AF phase with FM order in $ab$ planes and spin canting 
angle $\theta$ along the $c$ axis.
These phases are precursor states to the FM phase at large $\eta>0.2$,
see Fig. \ref{fig:cmf}(c).
}
\label{fig:spins}
\end{figure}

When $\eta$ is further increased within the $A$-AF phase, one finds a 
second magnetic transition, and two exotic phases are found: 
($i$) the striped-AF phase characterized by symmetry breaking between 
the $a$ and $b$ directions in the orbital and spin sectors for $E_z>0$,
see Fig. \ref{fig:spins}(a), and
($ii$) the canted-$A$-AF phase when the spins stay FM within the $ab$ 
planes, but rotate gradually from the AF to FM configuration along the 
$c$ bonds with increasing value of $\eta$, see Fig. \ref{fig:spins}(b).
Both phases are characterized by the spin angle --- the striped-AF 
phase by $\phi^a$ along the $a$ axis, and the canted-$A$-AF phase by
$\phi^c$ along the $c$ axis. The two phases shown in Fig. 
\ref{fig:spins} are quite different, as the CF term breaks the symmetry 
in the orbital space. 

\begin{figure}[t!]
    \includegraphics[width=8.0cm]{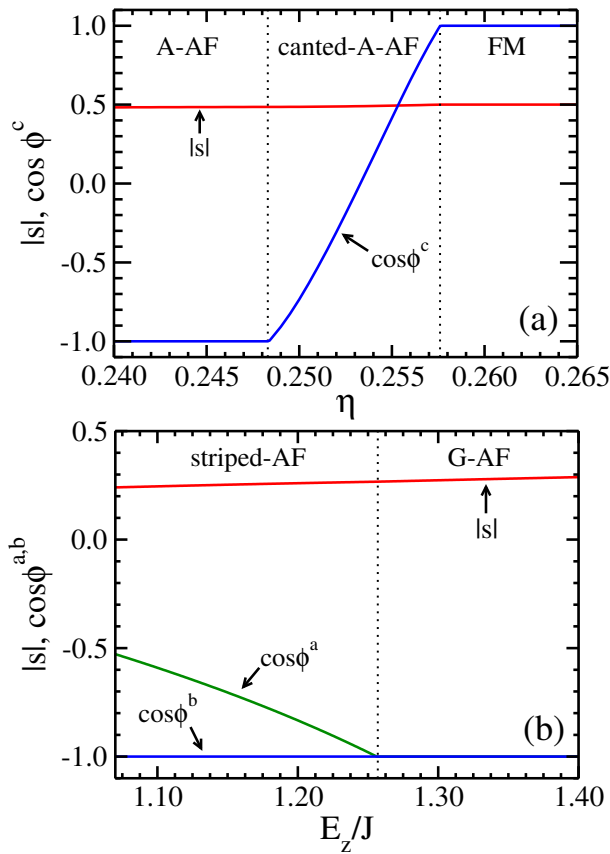}
\caption{
Spin order parameters $|s|$ (\ref{|s|}) and cosines of spin angles 
between two neighboring spins in the two exotic phases found in the 3D 
perovskite:
(a) evolution from the $A$-AF via the canted-$A$-AF to FM phase with 
increasing $\eta$ at $E_z=-0.5J$, and 
(b) evolution from the striped-AF to $G$-AF phase with increasing CF 
splitting $E_z/J$, found at $\eta=0.22$; here spin order $|s|$ is 
significantly reduced by quantum fluctuations.
}
\label{fig:evo}
\end{figure}

The transition to the FM order for negative CF splitting involves the 
intermediate canted-$A$-AF phase, shown for $E_{z}=-0.5J$ in Fig. 
\ref{fig:evo}(a). Here the spin order is already FM in the $ab$ planes, 
so it suffices to analyze the phase transition using a linear embedded 
cluster \cite{Brz13}. The evolution of the spin order is captured by 
two quantities: the spin canting angle $\phi^c$ along the $c$ axis, and 
the total magnetization $|s|$, defined as follows,
\begin{eqnarray}
\cos\theta&=&
\frac{1}{s^{2}}\left(s_{1}^{x}s_{2}^{x}+s_{1}^{z}s_{2}^{z}\right),
\label{cos}\\
|s|&\equiv&\sqrt{\left(s^{x}\right)^{2}+\left(s^{z}\right)^{2}}.
\label{|s|}
\end{eqnarray}
In the canted-$A$-AF phase $\cos\phi^c$ interpolates smoothly between
limits $\cos\phi^c=-1$ in the $A$-AF phase and $\cos\phi^c=1$ in the FM 
one. Figure \ref{fig:evo}(a) shows also that the spin order parameter 
$|s|$ is almost classical ($|s|\simeq 0.5$) in the $A$-AF phase as the 
quantum corrections are here rather low ---
the quantum spin fluctuations decrease further across the canted-$A$-AF 
phase and finally one finds the exact value $|s|=0.5$ in the FM phase.

The second example of the exotic spin order found in the 3D KK model 
considered here is the striped-AF phase which gradually turns into the 
$G$-AF N\'eel order when $E_{z}$ increases. In this case the plaquette
cluster in an $ab$ plane is more appropriate as it captures the changes 
of of the spin order here. The orbital order parameters $\{t^a,t^b\}$ 
(not shown) suggest the symmetry breaking within the $ab$ planes in the 
striped-AF phase which is restored at the transition to the $G$-AF 
phase. The spin order is given by four sublattices, two of them shown 
in Fig. \ref{fig:spins}(b), and the other two related to them by a spin 
inversion. 

The cosines of two angles between the neighboring spins along the $a$ 
and $b$ axis (\ref{cos}), $\phi^{a(b)}$, show that the AF order along 
the $b$ axis is independent of $E_z$, see Fig. \ref{fig:evo}(b). In the 
$G$-AF phase one finds $\cos\phi^{a(b)}=-1$, as expected for the uniform
antiferromagnet. The total magnetization $0.24<|s|<0.30$ 
(\ref{|s|}) is almost constant and increases monotonically with
increasing $E_z$. This demonstrates that the essential physics of the 
striped-AF phase is described by the spin angles, but also that the 
order parameter is much softened by spin fluctuations on the 
plaquettes within the $ab$ planes.

\section{Summary}
\label{sec:summa}

We have investigated the phase diagrams of the spin-orbital $d^9$
Kugel-Khomskii model by the mean-field and perturbative methods
for increasing system dimensionality: from the square lattice 
monolayer, via the bilayer to the cubic lattice. In each case we have 
found strong competition between different types of spin and orbital 
order, with entangled spin-orbital phases at the crossover from 
antiferromagnetic to ferromagnetic correlations in the intermediate 
regime of Hund's exchange. These phases stem from the quantum 
fluctuation of the ordered orbitals that couple to spins and produce 
novel types of spin bonds including non-trivial four-spin interactions.  

For the 2D monolayer the ortho-AF exotic phase was found, which is 
characterized by a non-collinear spin order where neighboring spins are 
perpendicular to each other and orbitals are strongly polarized in a 
FO$z$ configuration. Such phases are excluded in the single-site 
mean-field approach, and could not have been found before \cite{Cha08}. 
Both cluster mean-field and a more involved Entanglement Renormalization 
Ansatz (ERA) involving entanglement between clusters confirmed stability 
of the ortho-AF phase in between the AF and FM phases. 
On the other hand, the perturbative treatment in the orbital space was 
introduced to get effective spin Hamiltonian in the ortho-AF phase and
to provide the physical insight into mechanism stabilizing spins being 
at the same time AF on two interpenetrating sublattices and 
perpendicular for nearest neighbors when virtual orbital flips occur
and are accompanied by spin singlets. 

The case of a bilayer turned out to be rather special and different 
from both 2D and 3D case due to its strong tendency towards formation 
of interplanar singlets. Unlike the ortho-AF phase, the entangled 
phases found here are located in the intermediate coupling regime and 
are not triggered by a magnetic phase transition. Apart from this, 
the location of the long-range order phases and the plaquette valence 
bond phase is already quite similar to the 3D case where the ortho-AF 
phase is found again together with two additional phases with exotic 
spin order induced by orbital fluctuations. These striped-AF and 
canted-$A$-AF phases appear here at relatively high values of Hund's 
exchange in the vicinity of the FM phase, and essentially by the same 
mechanism as the ortho-AF phase in the 2D case. 

The most striking feature in the phase diagram of the 3D model is that 
for negative $E_z$, the transition from the fully AF to the fully FM 
configuration takes place gradually with growing Hund's coupling $\eta$, 
first the $ab$ planes become FM passing through the exotic ortho-AF 
phase, and next the remaining bonds along the $c$ axis become FM 
passing through the canted-$A$-AF exotic phase. In contrast, for large 
positive $E_z$ the AF and FM phases are connected directly by a 
discontinuous transition. The only exception occurs at intermediate 
$E_z>0$ where a striped-AF phase is stable. We argue that this last 
exotic phase with anisotropic AF order in the $ab$ planes is a 
positive-$E_z$ counterpart of the ortho-AF phase where the effective 
spin interaction of different orders compete with each other leading 
to frustration and anisotropy. 

Finally, the 2D monolayer system served us as a testing ground of the 
thermal stability of spin and orbital orders found in the KK model. 
We have found that orbital order is more robust in general and spin 
order melts first under increasing temperature, as for instance in 
LaMnO$_3$ \cite{Fei99}. Robust orbital interactions occur also in 
triangular lattice, as for instance in LiNiO$_2$ \cite{Rei05}. The 
most fragile is the exotic spin order found in ortho-AF phase which 
may be expected taking the energy scales of the effective spin 
couplings. The other factor that also strongly 
decreases the temperature of spin melting may be the orbital order 
which is incompatible with lattice geometry and strongly suppresses 
in-plane couplings, such as FO$z$ order found for negative 
crystal-field $E_z<0$. In contrast, the plaquette valence bond 
phase is robust type of order due to in-plane singlets. We believe 
that these general conclusions are valid for similar models as well.

\acknowledgments

We kindly acknowledge financial support by 
the Polish National Science Center (NCN) under Projects:
No. 2012/04/A/ST3/00331 (W.B. and A.M.O.)
and
No. 2013/09/B/ST3/01603 (J.D.).


\begin{thebibliography}{99}

\bibitem{Kug73} K.I. Kugel, D.I. Khomskii,
                   {\it JETP} \textbf{37}, 725 (1973).
                   
\bibitem{Kug82} K.I. Kugel, D.I. Khomskii,
                   {\it Sov. Phys. Usp.} \textbf{25}, 231 (1982).

\bibitem{Hfm}   J. van~den Brink, Z. Nussinov, A.M. Ole\'s,
                   in: {\it Introduction to Frustrated Magnetism: 
                   Materials, Experiments, Theory}, edited by C. Lacroix,
                   P. Mendels, F.~Mila (Springer, New York, 2011).

\bibitem{Kha05} G. Khaliullin,
                   {\it Prog. Theor. Phys. Suppl.\/} \textbf{160}, 155 (2005).

\bibitem{Ole05} A.M. Ole\'s, G. Khaliullin, P. Horsch, L.F. Feiner,
                   {\it Phys. Rev. B\/} \textbf{72}, 214431 (2005).

\bibitem{Ole09} A.M. Ole\'s,
                   {\it Acta Phys. Polon. A} \textbf{115}, 36 (2009).

\bibitem{Ole12} A.M. Ole\'s,
                   {\it J. Phys.: Condens. Matter\/} \textbf{24}, 313201 (2012).
                   
\bibitem{Fei97} L.F. Feiner, A.M. Ole\'s, J. Zaanen,
                   {\it Phys. Rev. Lett.\/} \textbf{78}, 2799 (1997);
                   {\it J. Phys. Condens. Matter\/} \textbf{10}, L555 (1998).
                   
\bibitem{Kha97} G. Khaliullin, V. Oudovenko,  
                   {\it Phys. Rev. B\/} \textbf{56}, R14243 (1997).

\bibitem{Tok06} Y. Tokura,
                   {\it Prog. Theor. Phys.\/} \textbf{69}, 797 (2006).
                   
\bibitem{Fuj10} J. Fujioka, T. Yasue, S. Miyasaka, Y. Yamasaki, T. Arima, 
                   H. Sagayama, T. Inami, K. Ishii, Y. Tokura,
                   {\it Phys. Rev. B\/} \textbf{82}, 144425 (2010).
                   
\bibitem{Goo06} J.-S. Zhou, J.B. Goodenough, 
                   {\it Phys.~Rev. Lett.\/} {\bf 96}, 247202 (2008).

\bibitem{Woh11} K. Wohlfeld, M. Daghofer, S. Nishimoto, G. Khaliullin, 
                   J. van den Brink,   
                   {\it Phys. Rev. Lett.\/} \textbf{107}, 147201 (2011).
                   
\bibitem{Woh13} K.~Wohlfeld, S. Nishimoto, M.W. Haverkort, J. van den Brink,   
                   {\it Phys. Rev. B\/}      \textbf{88}, 195138 (2013).
                   
\bibitem{Ole06} A.M. Ole\'s, P. Horsch, L.F. Feiner, G. Khaliullin,
                   {\it Phys. Rev. Lett.\/} \textbf{96}, 147205 (2006).

\bibitem{Che07} Y. Chen, Z.D. Wang, Q.Y. Li, F.C. Zhang,
                   {\it Phys. Rev. B\/} \textbf{75}, 195113 (2007).
                   
\bibitem{Nor08} B. Normand, A.M. Ole\'s,
                   {\it Phys. Rev. B\/} \textbf{78}, 094427 (2008).
                   
\bibitem{Cha11} J. Chaloupka, A.M. Ole\'s,
                   {\it Phys. Rev. B\/} \textbf{83}, 094406 (2011).

\bibitem{Her11} A. Herzog, P. Horsch, A.M. Ole\'s, J. Sirker,
                   {\it Phys. Rev. B\/}  \textbf{83}, 245130 (2011).
                   
\bibitem{You12} W.-L You, A.M. Ole\'s, P. Horsch,
                   {\it Phys. Rev. B\/} \textbf{86}, 094412 (2012).
 
\bibitem{Lun12} R. Lundgren, V. Chua, G.A. Fiete, 
                   {\it Phys. Rev. B\/} \textbf{86}, 224422 (2012).

\bibitem{Ulr03} C. Ulrich, G. Khaliullin, J. Sirker, M. Reehuis, M. Ohl, 
                   S. Miyasaka, Y. Tokura, B. Keimer, 
                   {\it Phys. Rev. Lett.\/} \textbf{91}, 257202 (2003).
                   
\bibitem{Hor03} P. Horsch, G. Khaliullin, A.M. Ole\'s,    
                   {\it Phys. Rev. Lett.\/} \textbf{91}, 257203 (2003).
                
\bibitem{Sir08} J.~Sirker, A. Herzog, A.M. Ole\'s, P. Horsch,
                   {\it Phys. Rev. Lett.\/} \textbf{101}, 157204 (2008).
                   
\bibitem{Hor08} P. Horsch, A.M. Ole\'s, L.F. Feiner, G. Khaliullin, 
                   {\it Phys.~Rev. Lett.\/} {\bf 100}, 167205 (2008).

\bibitem{Nor09} Bruce Normand,
                   {\it Cont. Phys.\/} \textbf{50}, 533 (2009).

\bibitem{vdB99} J. van~den Brink, P. Horsch, F. Mack, A.M. Ole\'s, 
                   Phys. Rev. B \textbf{59}, 6795 (1999).
                   
\bibitem{vdB04} J. van den Brink, 
                   New J. Phys. \textbf{6}, 201 (2004).
                   
\bibitem{Ryn10} A. van Rynbach, S. Todo, S. Trebst, 
                   Phys. Rev. Lett. \textbf{105}, 146402 (2010).  

\bibitem{Mil05} J. Dornier, F. Becca, F. Mila,
                   {\it Phys. Rev. B\/} \textbf{72}, 024448 (2005).
                   
\bibitem{Cin10} L. Cincio, J. Dziarmaga, A.M.~Ole\'s,
                   {\it Phys. Rev. B\/} \textbf{82}, 104416 (2010).
                   
\bibitem{Tro10} F. Trousselet, A.M. Ole\'s, P. Horsch,
                   {\it Europhys. Lett.\/} \textbf{91},  40005 (2010);
                   {\it Phys. Rev. B\/}    \textbf{86}, 134412 (2012).
                   
\bibitem{Brz10} W. Brzezicki, A.M. Ole\'s,
                   {\it Phys. Rev. B\/} \textbf{82}, 060401 (2010);
                   {\it Phys. Rev. B\/} \textbf{87}, 214421 (2013).
                  
\bibitem{BrzDa} W. Brzezicki, M. Daghofer, A.M. Ole\'s,
                   {\it Phys. Rev. B\/} \textbf{89}, 014431 (2014).
                  
\bibitem{Fri99} B. Frischmuth, F. Mila, M. Troyer,                
                   {\it Phys. Rev. Lett.\/} \textbf{82}, 835 (1999).
                   
\bibitem{Ole07} A.M. Ole\'s, P. Horsch, G. Khaliullin,                   
                   {\it Phys. Stat. Solidi B\/} \textbf{244}, 2378 (2007).
                   
\bibitem{Kum13} B. Kumar, 
                   {\it Phys. Rev. B\/} \textbf{87}, 195105 (2013).
                   
\bibitem{Brz14} W. Brzezicki, J. Dziarmaga, A.M. Ole\'s,
                   {\it Phys. Rev. Lett.\/} \textbf{112}, 117204 (2014).                  

\bibitem{Ten93} D.A. Tennant, T.G. Perring, R.A. Cowley, S.E. Nagler,
                   {\it Phys. Rev. Lett.\/} \textbf{70}, 4003 (1993).
                   
\bibitem{Dei08} J. Deisenhofer, I. Leonov, M.V. Eremin, Ch. Kant, P. Ghigna, F. Mayr, 
                   V.V. Iglamov, V.I. Anisimov, D. van der Marel,
                   {\it Phys. Rev. Lett.\/} \textbf{101}, 157406 (2008).
                   
\bibitem{Lak13} B. Lake, D.A. Tennant, J.-S. Caux, T. Barthel, U. Schollw\"ock, 
                   S.E. Nagler, C. D. Frost,
                   {\it Phys. Rev. Lett.\/} \textbf{111}, 137205 (2013).  
                   
\bibitem{Fei99} L.F. Feiner, A.M. Ole\'s, 
                   {\it Phys. Rev. B\/} \textbf{59}, 3295 (1999).
                   
\bibitem{Jef92} J.H. Jefferson, H. Eskes, L.F. Feiner,
                   {\it Phys. Rev. B\/} \textbf{45}, 7959 (1992).
                   
\bibitem{Mos02} M. Mostovoy, D.I. Khomskii,
                   {\it Phys. Rev. Lett.\/} \textbf{89}, 227203 (2002).
                   
\bibitem{Ole00} A.M. Ole\'s, L.F. Feiner, J. Zaanen,
                   {\it Phys. Rev. B\/} \textbf{61}, 6257 (2000).

\bibitem{Ole83} A. M. Ole\'s,
                   {\it Phys. Rev. B\/} \textbf{28}, 327 (1983).

\bibitem{Brz12} W. Brzezicki, J. Dziarmaga, A.M. Ole\'s,
                   {\it Phys. Rev. Lett.\/} \textbf{109}, 237201 (2012).
                   
\bibitem{Brz11} W. Brzezicki, A.M. Ole\'s,
                   {\it Phys. Rev. B\/} \textbf{83}, 214408 (2011).

\bibitem{Brz13} W. Brzezicki, J. Dziarmaga, A.M. Ole\'s,
                   {\it Phys. Rev. B\/}     \textbf{87}, 064407 (2013).
                   
\bibitem{Cha77} K.A. Chao, J. Spa\l{}ek, A.M. Ole\'s,
                   {\it J. Phys. C\/}       \textbf{10}, L271 (1977).
                   
\bibitem{Cha08} J. Chaloupka, G. Khaliullin,       
                   {\it Phys. Rev. Lett.\/} \textbf{100}, 016404 (2008).
                   
\bibitem{Rei05} A.J.W. Reitsma, L.F. Feiner, A.M. Ole\'s,
                   {\it New J. Phys.\/} \textbf{7}, 121 (2005).
                   
\end{thebibliography}
\end{document}